\documentclass[11pt]{article}

\begin{document}
\title{\Large \bf  Causal Cones, Cone Preserving Transformations and Causal Structure in Special and General Theory of Relativity* }
\date{}
\maketitle{\noindent\small {{Sujatha \ Janardhan$\ ^{1}$ and  R \ V \
Saraykar$\ ^{2}$}}  \\$\ ^{1}$ {Department of Mathematics,
St.Francis De Sales College,  Nagpur-440 006. \\ e-mail :
sujata\_jana@yahoo.com,  \  Ph : 91712-2571896
\\ $\ ^{2}$ {Department of Mathematics, R\ T\ M Nagpur University, Nagpur-440
033. \\ e-mail :  ravindra.saraykar@gmail.com}}} \\

\noindent \small (*To appear in \emph{Gravitation and Cosmology}, vol. 19, issue 1 (2013)). 

\vspace{5mm}

\noindent\textbf{Abstract} \ We present a short review of geometric and algebraic approach to
 causal cones and describe cone preserving transformations and their relationship with causal
 structure related to special and general theory of relativity. We describe Lie groups,
 especially matrix Lie groups, homogeneous and symmetric spaces  and  causal cones
 and certain implications of these concepts in special and general theory of relativity related
 to causal structure and topology of space-time. We  compare and  contrast the results on causal
 relations with those in the literature for general space-times and compare these relations
 with K-causal maps. We also describe causal orientations and their implications for space-time
 topology and discuss some more topologies on space-time which arise as an application of domain theory.

\section{Introduction}

The notion of causal order is a basic concept in physics and in
the theory of relativity in particular. A space time metric
determines causal order and causal cone structure. Alexandrov
\cite{alex1,alex2} proved that a causal order can determine a
topology of space-time called Alexandrov topology which, as is now
well known, coincides with manifold topology if the space time is
strongly causal. The books by Hawking-Ellis, Wald and Joshi
\cite{H&E,W,Jo} give a detailed treatment of causal structure of
space-time. However, while general relativity employs a Lorentzian
metric, all genuine approaches to quantum gravity are free of
space-time metric. Hence the question arises whether there exists
a structure which gets some features of causal cones (light cones)
in a purely topological or order-theoretic manner. Motivated by
the requirement on suitable structures for a theory of quantum
gravity, new notions of causal structures and cone structures were
deployed on a space-time.\\
The order theoretic structures, namely causal sets have been
extensively used by Sorkin and his co-workers in developing a new
approach to quantum gravity \cite{DHS}. As a part of this program,
Sorkin and Woolgar \cite{SW} introduced a relation called
K - causality and proved interesting results by making use of
Vietoris topology. Based on this work and other recent work, S.
Janardhan and R.V.Saraykar \cite{JS1,JS2} and E.Minguzzi \cite{Mi1,Mi2}
proved many interesting results. Especially after good deal of
effort, Minguzzi \cite{Mi1} proved that K - causality condition is
equivalent to \emph{stably causal} condition.

\noindent More recently, K.Martin and Panangaden \cite{MP} making use of domain
theory, a branch of theoretical computer science, proved
fascinating results in the causal structure theory of space-time.
The remarkable fact about their work is that only \emph{order} is
needed to develop the theory and topology is an outcome of the
\emph{order}. In addition to this consequence, there are abstract
approaches, algebraic as well as geometric to the theory of cones
and cone preserving mappings. Use of quasi-order (a relation which
is reflexive and transitive) and partial order is made in defining
the cone structure. Such structures and partial orderings are used
in the optimization problems \cite{Tru}, game theory and decision making
etc \cite{Lu}. The interplay between ideas from theoretical computer
science and causal structure of space-time is becoming more
evident in the recent works \cite{Kre1,Kre2}.

\noindent Keeping in view these developments, in this paper, we
present a short review of geometric and algebraic approach to
causal cones and describe cone preserving transformations and
their relationship with causal structure. We also describe certain
implications of these concepts in special and general theory of
relativity related to causal structure and topology of space-time.

\noindent Thus in section 2, we begin with describing Lie groups, especially
matrix Lie groups, homogeneous spaces and then causal cones. We
give an algebraic description of cones by using quasi-order.
Furthermore, we describe cone preserving transformations. These
maps are generalizations of causal maps related to causal
structure of space-time which we shall describe in section 3. We
then describe explicitly Minkowski space as an illustration of
these concepts and note that some of the space-time models in
general theory of relativity can be described as homogeneous
spaces.

\noindent In section 3, we describe causal structure of space time,
causality conditions, K-causality and hierarchy among these
conditions in the light of recent work of Minguzzi and Sanchez
\cite{MiS}. We also describe geometric structure of causal
group, a group of transformations preserving causal structures or
a group of causal maps on a space-time.

\noindent In section 4, we describe causal orientations and their
implications for space-time topology. We find a parallel between
these concepts and concepts developed by Martin and Panangaden
\cite{MP} to describe topology of space time, especially a globally
hyperbolic one.Finally we discuss some more topologies on
space-time which arise as an application of domain theory.
\\ We end the paper with concluding remarks.


\section{ Causal Cones and cone preserving transformations}

To begin with, we describe Lie groups, matrix Lie groups,
homogeneous and symmetric spaces and state some results about
them. These will be used in the discussion on causal cones. We
refer to the books  \cite{GIL,SH,JAW} for more details.

\noindent\textbf{Definition :} \ Lie groups and matrix Lie groups:

\noindent \textbf{Lie group}: A finite dimensional manifold $\ G$ is
called a Lie group if $\ G$ is a group such that the group operations,
composition and inverse are compatible with the differential
structure on $\ G$. This means that the mappings\\
$\ G \times G  \rightarrow G : (x,y) \mapsto  x.y $ and \\
$\ G \rightarrow G : x \mapsto x^{-1} $ \\
are $\ C^{\infty}$ as mappings from one manifold to other.

\noindent The n-dimensional real Euclidean space $\ R^{n}$, n-dimensional
complex Euclidean space $\ C^{n}$, unit sphere $\ S^{1}$ in $\
R^{2}$, the set of all $\ n \times n $ real matrices $\ M(n,R)$
and the set of all $\ n \times n $ complex matrices $\ M(n,C)$ are
the simplest examples of Lie groups. $\ M(n,R)$ (and $\  M(n,C) $)
have subsets which are Lie groups in their own right. These Lie
groups are called \emph{matrix Lie groups}. They are important
because most of the Lie groups appearing in physical sciences such
as classical and quantum mechanics,  theory of relativity -
special and general, particle physics etc are matrix Lie groups.
We describe some of them here, which will be used later in this
article.\\
{$\ \textbf{Gl(n,R)}$} : General linear group of $\ n \times n $
real invertible matrices. It is a Lie group and topologically an
open subset of $\ M(n,R)$. Its dimension is $\ n^{2} $.\\
$\ \textbf{Sl(n,R)}$ : Special linear group of $\ n \times n $
real invertible matrices with determinant +1. It is a closed
subgroup of $\ Gl(n,R)$ and a Lie group in its own right, with
dimension $\ n^{2}-1 $.\\
$\ \textbf{O(n)}$: Group of all $\ n \times n $ real orthogonal
matrices. It is called an orthogonal group. It is a Lie group of
dimension $\ \frac{n(n-1)}{2}$.\\
$\ \textbf{SO(n)}$: Special orthogonal group- It is a connected
component of $\ O(n)$ containing the identity I and also a closed
(compact) subgroup of $\ O(n) $ consisting of real orthogonal
matrices with determinant +1. In particular $\ SO(2)$  is
isomorphic to $\ S^{1} $.\\
The corresponding Lie groups which are subsets of $\  M(n,C)$  are
$\ GL(n,C)$, $\ SL(n,C)$, $\ U(n)$ and $\ SU(n)$ respectively,
where \emph{orthogonal} is replaced by \emph{unitary}. $\ SU(n)$
is a compact subgroup of $\ GL(n,C)$. For n=2, it can be proved
that $\ SU(2)$ is  isomorphic to $\ S^{3}$, the unit sphere in $\
R^{4}$. Thus $\ S^{3}$ is a Lie group.[However for topological
reasons, $\ S^{2}$ is not a Lie group, though it is
$\ C^{\infty}$ -differentiable manifold]\\
$\ \textbf{O(p,q) and  SO(p,q)}$ : Let p and q be positive
integers such that $\ p+q = n$. Consider the quadratic form $\
Q(x_{1},x_{2}...x_{n})$ given by\\
$\ Q= x_{1}^{2} + x_{2}^{2}+... x_{p}^{2}-x_{p+1}^{2}-x_{p+2}^{2}...-x_{n}^{2}.$\\
The set of all $\ n \times n$  real matrices which preserve this
quadratic form Q is denoted by $\ O(p,q)$ and a subset of $\
O(p,q)$ consisting of those matrices of $\ O(p,q)$ whose
determinant is +1, is denoted by $\ SO(p,q)$. Both $\ O(p,q)$ and
$\ SO(p,q)$ are Lie groups. Here preserving quadratic form Q means
the
following:\\
Consider standard inner product $\ \eta$ on $\ R^{p+q} = R^{n}
$given by the diagonal matrix: \\
$\ \eta = diag(1,1 \ldots 1,-1,-1 \ldots -1)$, (1 appearing p times).\\
Then $\ \eta$ gives the above quadratic form $\
Q(x_{1},x_{2},...,x_{n})$,\\
i.e.$\  X \eta X^{T} = Q(x_{1},x_{2},...,x_{n})$ where $\ X=
[x_{1},x_{2},...,x_{n}]$. $\ n \times n$ matrix A is said to
preserve the quadratic form Q if $\ A^{T} \eta A = \eta $.

\noindent $\ O(p,q)$ is called \emph{indefinite orthogonal group} and $\
SO(p,q)$ is called \emph{indefinite special orthogonal group}.
Dimension of $\ O(p,q)$ is $\ \frac{n(n-1)}{2}$.\\
Assuming both p and q are nonzero, neither of the groups $\
O(p,q)$ or $\ SO(p,q)$ are connected. They have respectively four
and two connected components. The identity component of $\ O(p,q)$
is denoted by $\ SO_{o}(p,q)$ and can be identified with the set
of elements in $\ SO(p,q)$ which preserves both orientations.

\noindent In particular $\ O(1,3)$ is the Lorentz group, the group of all
Lorentz transformations, which is of central importance for
electromagnetism and special theory of relativity. $\ U(p,q)$ and
$\ SU(p,q)$ are defined similarly. For more details, we refer the
reader to \cite{GIL,Kn}

\noindent We now define Homogeneous spaces and discuss some of their
properties:

\noindent\textbf{Definition :} \ We say that a Lie group $\ G$ is
\emph{represented as a Lie group of transformations of a
$\ C ^{\infty}$ manifold M (or has a left (Lie)- action on M)} if to
each  $\ g \in G $, there is associated a diffeomorphism from $\ M
$ to itself: \ $\ x \mapsto \psi_{g}(x) , x \in M $ such that $\
\psi_{gh} = \psi_{g} \psi_{h}$ for all $\ g, h \in G $ and $\
\psi_{e} = Id.$, Identity map of $\ M $, and if further-more $\
\psi_{g}(x) $ depends smoothly on the arguments $\ g $, $\ x $.
i.e. the map $\ (g, x) \mapsto \psi_{g}(x) $ is a smooth map from
$\ G \times M \rightarrow M $.

\noindent The Lie group $\ G$ is said to have a \emph{right action} on M if the
above definition is valid with the  property $\ \psi_{g} \psi_{h}
= \psi_{gh} $ replaced by $\ \psi_{g} \psi_{h} =  \psi_{hg} $.

\noindent If $\ G$ is any of the matrix Lie groups\\
$\ GL(n,R), O(n,R), O(p,q)$ or  \\ $\  GL(n, C), U(n), U(p,q)$ (where $\ p + q = n)$, then $\ G$
acts in the obvious way on the manifold $\ R^{n} $ or $\ R^{2n} =
$\ C $^{n} $. In these cases, the elements of $\ G$ act as linear
transformations.

\noindent The action of a group $\ G$ is said to be \emph{transitive} if for
every two points $\ x, y $ of $\ M $, there exists an element $\ g
\in G $ such that $\ \psi_{g}(x) = y $.\\
\noindent\textbf{Definition :} \ A manifold on which a Lie group
acts transitively is called a \emph{homogeneous space} of the Lie
group.

\noindent In particular, any Lie group $\ G$ is a homogeneous space for itself
under the action of left multiplication. Here $\ G$ is called the
\emph{Principal left homogenous space} (of itself).  Similarly the
action $\ \psi_{g}(h) =  h g^{-1} $ makes $\ G$ into its own
\emph{Principal right homogeneous space.}

\noindent Let $\ x $ be any point of a homogeneous space of a Lie group $\ G$.
The \emph{isotropy group }(or stationary group) $\ H_{x} $ of the
point $\ x $ is the stabilizer of $\ x $ under the action of $\ G$ :
$\ H_{x} = \{ g \in G /  \psi_{g}(x) = x \} $. \\
We have the following lemma.

\noindent\textbf{Lemma :} All isotropy groups $\ H_{x}$  of
points $\ x $ of a homogeneous space are isomorphic.

\noindent\textbf{Proof :} Let $\ x $, y be any two points of the
homogeneous space. Let $\ g \in G $ be such that $\ \psi_{g}(x) =
y $. Then the map $\ H_{x} \rightarrow H_{y} $ defined by $\ h
\mapsto ghg^{-1} $  is an isomorphism. ( Here we have assumed the
left action).

\noindent We thus denote simply by H, the isotropy group of some (and hence
of every element modulo isomorphism)  element of $\ M $ on which $\ G$
acts on the left.\\
We now have the following theorem.

\noindent \textbf{Theorem :} There is a one- one
correspondence between the points of a homogeneous space $\ M $ of
the Lie group $\ G$, and the left cosets gH of H in $\ G$, where H is the
isotropy group and $\ G$ is assumed to act on the left.

\noindent\textbf{Proof: } Let $\ x_{0}$  be any point of the
manifold $\ M $. Then with each left coset $\ gH_{x_{0}}$  we
associate the point $\ \psi_{g}(x_{0})$  of $\ M $. Then this
correspondence is well- defined, i.e. independent of the choice of
representative of the coset, one - one and onto.

\noindent It can be shown under certain general conditions that the isotropy
group H is a closed sub group of $\ G$ , and the set $\ G/H $ with the
natural quotient topology can be given a unique (real) analytic
manifold structure such that $\ G$ is a Lie transformation group of $\
G/H $. Thus $\ M  \approx  G/H $.

\noindent\textbf{Examples of homogeneous spaces are:}

\noindent\textbf{1. Stiefel manifolds : } \ For each $\ n, k  ( k
\leq n)$, the Stiefel manifold $\ V_{n,k} $ has as its points all
orthonormal frames $\ x = (e_{1},e_{2}...,e_{k})$ of $\ k $
vectors in Euclidean $\ n $-space i.e. ordered sequences of $\ k $
orthonormal vectors in $\ R^{n} $. Then $\ V_{n,k} $ is embeddable
as a non- singular surface of dimension $\ nk - k (k+1)/2 $ in $\
R^{nk}$ and can be visualized as $\ SO(n) / SO(n-k) $. In
particular we have $\ V_{n,n} \cong O(n),  V_{n,n-1} \cong SO(n),
\  V_{n,1} \cong S^{n-1}$.

\noindent\textbf{ 2. Grassmannian manifolds :} \ The points of the
Grassmannian manifold $\ G_{n,k}$, are by definition, the $\ k $-
dimensional planes passing through the origin of $\ n
$-dimensional Euclidean space. This is a smooth manifold and it is
given by \\ $\ G_{n,k} \cong  O(n)/ O(k) \times O(n-k)$.
\\ We now define  symmetric spaces.

\noindent\textbf{Definition :} A simply connected manifold $\ M $
with a metric $\ g_{ab}$ defined on it, is called a
\emph{symmetric space} (\emph{symmetric manifold}) if for every
point $\ x $ of $\ M $, there exists an isometry (motion) $\ s_{x}
: M \rightarrow M $ with the properties that $\ x $ is an isolated
fixed point of it, and that the induced map on the tangent space
at $\ x $ reflects ( reverses ) every tangent vector at $\ x $
i.e. $\ \xi \mapsto  - \xi $. Such an isometry is called \emph{a
symmetry of M of the point $\ x $}.
\\ For every symmetric space, covariant derivative of Riemann
curvature tensor vanishes.
\\ For a homogeneous symmetric manifold $\ M $, let $\ G$ be the Lie
group of all isometries of $\ M $ and let H be the isotropy group
of $\ M $ with respect to left action of $\ G$ on $\ M $. Then , as we
have seen above, $\ M $ can be identified with $\ G/H $, the set of
left cosets of H in $\ G$. As examples of such spaces in general
relatively, we have the following space-times:

\noindent\textbf{Space of constant curvature with isotropy group
$\ H = SO(1,3)$}:

\noindent 1. Minkowski space $\ R^{4} $. \\
2. The de Sitter space \\
$\ S_{+} = SO(1,4) / SO(1,3)$. Here $\
S_{+}$ is homeomorphic to $\ R \times S^{3}$ and  the curvature
tensor $\ R $ is the identity operator on the space of bivectors $\
\Lambda^{2}(R^{4}) , R = Id $. \\
3. The anti- de Sitter space \\
$\ S = SO( 2,3) / SO(1,3)$. This
space is homeomorphic to $\ S^{1} \times R^{3}$ and  its universal
covering space is homeomorphic to $\ R^{4}$. Here curvature tensor
$\ R = - Id $.

\noindent Another example of symmetric space-time is the symmetric
space $\ M_{t} $ of plane waves. For these spaces the isotropy
group is abelian, and the isometry group   is soluble
(\emph{solvable}). (A group $\ G$ is called \emph{solvable} if it has
a finite chain of normal subgroups $\ \{e\} < G_{1} <...< G_{r }=
G $, beginning with the identity subgroup and ending with $\ G$, all
of whose factors $\ G_{i +1}/G_{i}$ are abelian). In terms of
suitable coordinates, the metric has the form \\ $\ ds^{2} =
2dx_{1} \ dx_{4}+ [ (cos \ t) x_{2}^{2} + (sin \ t)x_{3}^{2} ] \
dx_{4}^{2} + \ dx_{2}^{2} + \ dx_{3}^{2}$, $\ cos \ t \geq  \ sin
\ t $. The curvature  tensor is constant (refer \cite{JAW}).

\noindent G\"{o}del universe \cite{H&E} is also an example of
a homogeneous space but it is not a physically reasonable model
since it contains closed time like curve through every point.
We now turn our attention to Causal cones and cone preserving transformations.\\
We note that all genuine approaches to quantum gravity are free
of space-time metric while general relativity employs a Lorentzian
space-time metric. Hence, the question arises whether there exists
a structure which gets some features of light cones in a purely
topological manner. Motivated by the requirements on suitable
structures for a theory of quantum gravity,  new notions of causal
structure and cone structures were developed on a space-time $\ M $.
Here we describe these notions.\\
The definition of \emph{causal cone} is given as follows:\\
Let $\ M $ be a finite dimensional real Euclidean vector (linear)
space with inner product $\ < , > $.  Let $\ R^{+} $ be the set of
positive real numbers and $\ R^{+}_{0} = R^{+} \cup \{0\}$ . A
subset $\ C $ of $\ M $  is a \emph{cone} if $\ R^{+} C \subset C $ and
is a \emph{convex cone} if $\ C $, in addition, is a convex subset of
$\ M $. This means, if $\ x,y \in C $ and $\ \lambda \in [0,1] $,
then $\ \lambda x + (1-\lambda) y \in C $. In other words, $\ C $ is a
convex cone if and only if for all $\ x,y \in C $ and $\ \lambda,
\mu \in R^{+}, \lambda x + \mu y \in C.$ We call cone $\ C $ as
\emph{non- trivial }if $\ C \neq -C $. If $\ C $ is non-trivial, then
$\ C \neq \{0\} $ and $\ C \neq M $ .\\
We use the following notations:\\
i.  $\ M^{c} = C \cap - C $ \\
ii. $\ <C> = C - C = \{ x - y / \  x,y \in C \}$ \\
iii. $\ C^{*} = \{ x \in M / \forall \ y \in C, (x,y) \geq 0 \} $\\
Then $\ M^{c} $ and $\ <C> $ are vector spaces. They are called
the \emph{edge} and the \emph{span} of $\ C $. The set $\ C^{*}$  is a
closed convex cone called the \emph{dual cone} of $\ C $. This
definition  coincides with the usual definition of the dual space
$\ M^{*} $ of $\ M $ by using inner product ( , ). If $\ C $ is a
closed convex cone, we have $\ C^{**} = C $,  and $\ (C^{*} \cap -
C^{*} ) = < C > ^{\bot} ,$ where for $\ U \subset M , \ U^{\bot} =
\{ y \in M / \forall u \in U, (u, y ) = 0\} $.

\noindent\textbf{Definition :} Let $\ C $ be a convex cone in $\ M $.
Then $\ C $ is called \emph{generating} if $\ < C > = M $. $\ C $ is called
\emph{pointed} if there exists a $\ y \in M $ such that for all $\
x \in C - \{0 \} $, we have $\ (x,y) >  0 $. If $\ C $ is closed , it
is called \emph{proper} if $\ M^{c} = \{0\}$. $\ C $ is called
\emph{regular} if it is generating and proper. Finally, $\ C $ is
called \emph{self-dual}, if $\ C^{*} = C $.\\
If $\ M $ is an ordered linear space, the Clifford's theorem
\cite{Fu} states that $\ M $ is directed if and only if $\ C $ is generating.\\
The set of interior points of $\ C $ is denoted by $\ C^{o} $ or $\ int
(C)$. The interior of $\ C $ in its linear span $\ <C> $ is called the
\emph{algebraic interior} of $\ C $ and is denoted by  \ alg int($\ C $).\\
Let $\ S \subset M$. Then the closed convex cone generated by $\ S $ is
denoted by \emph{Cone($\ S $)}:\\
Cone($\ S $) = \\
closure of $\ \{ \displaystyle\sum_{finite} r_{s} s / s
\in S , r_{s} \geq 0 \} $.

\noindent If $\ C $ is a closed convex cone, then its interior $\ C^{o} $ is an
open convex cone. If $ \Omega $ is an open convex cone, then its
closure  $\ \overline{\Omega} = cl(\Omega)  $ is a closed convex
cone. For an open convex cone, we define the dual cone by \\
$\ \Omega^{*}  = \{ x \in v /  \forall  y \in \overline{\Omega} -
\{0\}  \ (x, y ) > 0 \} \ = int( \overline{\Omega^{*}})$ . \\
If $\ \overline{\Omega} $ is proper, we have $\ \Omega^{**} =
\Omega$ \\
We  now have the  following results: \ ( cf   \cite{HN})

\noindent\textbf{Proposition :} Let $\ C $ be a closed convex cone
in $\ M $. Then
the following statements are equivalent:\\
i.  $\ C^{o} $ is nonempty \\
ii. $\ C $ contains a basis of $\ M $.\\
iii. $\ < C > =  M $

\noindent\textbf{Proposition :} Let $\ C $ be a  nonempty closed
convex cone
in $\ M $. Then the following properties are equivalent :\\
i.  $\ C $ is pointed \\
ii. $\ C $ is proper\\
iii.    int ($\ C^{*}) \neq \phi $\\
As a  consequence, we have

\noindent\textbf{Corollary :} Let $\ C $ be a closed convex cone.
Then $\ C $ is proper if  and only if $\ C^{*}$ is generating.

\noindent\textbf{Corollary :} Let $\ C $ be a convex cone in $\ M $.
Then $\ C \in Cone (M) $ if and only if $\ C^{*} \in Cone (M) $.
Here $\ Cone(M)$ is the set of all closed regular convex cones in
$\ M $.

\noindent To proceed further along these lines, we need to make ourselves
familiar with more terminology and notations. The linear
automorphism group of a convex cone is defined as follows:\\
Aut ($\ C $) = $\{ a \in GL (M) / \alpha(C) = C \} $. GL ($\ M $) is
the group of invertible linear transformations of $\ M $. If $\ C $ is
open or closed, Aut ($\ C $) is closed in GL ($\ M $). In particular
Aut($\ C $) is a linear Lie group.

\noindent\textbf{Definition :} Let $\ G$ be a group acting linearly on
$\ M $. Then a cone $\ C \in M $ is called \emph{$\ G$- invariant} if
$\ G.C = C $. We denote the set of invariant regular cones in $\ M $ by
$\ Cone_{G}(M)$. A convex cone $\ C $ is called \emph{homogeneous} if
Aut ($\ C $) acts transitively on $\ C $.

\noindent For $\ C \in Cone_{G} (M)$, we have Aut ($\ C $) = Aut $\
(C^{o})$ and $\ C = \partial C \cup C^{o} = (C-C^{o}) \cup C^{o} $
is a decomposition of $\ C $ into Aut ($\ C $) - invariant subsets. In
particular a non-trivial closed \emph{regular} cone can never be
homogeneous. \noindent We now state the following theorem:

\noindent\textbf{Theorem :} Let $\ G$ be a Lie group acting
linearly on the Euclidean vector space $\ M $ and $\ C \in
Cone_{G}(M)$. Then the stabilizer in $\ G$ of a point in $\  C^{o}$ is
compact.

\noindent In the abstract mathematical setting, cones  are described
using quasi-order relation  \cite{CH} as follows:

\noindent Let $\ M \neq 0 $ be a set and * be a mapping of $\ M
\times M $ into $\ \textsl{P}^{*}(M)$ (the set of all non-empty
subsets of $\ M $). The pair $\ ( M, * ) $ is called a
hypergroupoid. For $\ A, B \in \textsl{P}^{*}(M)$, we define $\ A*
B = \bigcup \{ a*b : a \in A, b \in B \}$.

\noindent A hypergroupoid $\ ( M,*)$ is called a hypergroup, if $\
(a*b)*c = a* (b*c) $ for all $\ a,b,c \in M$, and the reproduction
axiom, $\ a * M = M = M * a $, for any $ a \in M $,  is satisfied.

\noindent For a binary relation R on A and $\ a \in A $ denote $\ U_{R}(a) =
\{ b \in A/ <a , b> \in R \} $. A binary relation Q on a set A is
called quasiorder if it is reflexive and transitive. The set $\
U_{Q}(a)$  is called a cone of $\ a $. In the case when a
quasiorder Q is an equivalence,  $\ U_{Q}(A) = \{ x \in M /
\exists \ y \in A,  <x, y> \in \ Q \}$ for any $\ A \subseteq M $.
Analogously, for $\ B \subseteq A $ we set $\ U_{Q}(B) = \bigcap
\{ U_{Q}(a)/ a \in B \}$.

\noindent In the light of this definition, we shall observe in section 3 that causal cones
and K- causal cones fall in this category since causal relation $\ < $
and K-causal relation $\ \prec $ are reflexive and transitive.\\
In the literature, ( see for example \cite{Le1,Le2,ZK}), cone
preserving mappings are defined as follows:

\noindent Let $\ \emph{\textbf{A}} = ( A, R ) $ and $\ \textbf{\emph{B}} = (B, S) $ be
quasi-ordered sets. A mapping $\ h: A \rightarrow B $ is called \emph{cone preserving}
if $\ h( U_{R}(a)) = U_{S}(h(a))$ for each $\ a \in A. $

\noindent To illustrate the concepts described above,  we consider the
following example:

\noindent\textbf{Example  of a Forward Light cone in Minkowski
space :}

\noindent \textbf{Note:} In the paper by Gheorghe and Mihul \cite{GM},
forward light cone is called \emph{`positive cone'}and is defined as follows:\\
Let M be a n-dimensional  real linear space. A causal relation of
M is a partial ordering relation $\ \geq $ of M with regard to
which M is \textit{directed }, i.e. for any $\ x, y \in M $ there
is $\ z \in M $ so that $\ z \geq x, z \geq y $. Then the positive
cone is defined as $\ C = \{x/ x \in M; x \geq 0 \}$

\noindent Let $\ p $ and $\ q $ be two positive integers and $\ n = p + q $. Let $\ M =
R^{n} $. We write elements of $\ M $ as $\ v = \left(
\begin{array}{r}
x \\
y
\end{array}
\right) $  with $\ x \in R^{p}$ and $\ y \in R^{q} $. For $\ p $ = 1 ,
$\ x $ is a real number.\\
We write projections $\ p_{r_{1}} $ and $\ p_{r_{2}}$  as $\
p_{r_{1}}(v)  = x $ and  $\ p_{r_{2}}(v)  = y $.

\noindent As discussed earlier, connected component of identity in $\ O(p,q)
$ denoted by  $\ O(p,q)_{o} = SO_{0}(p,q)= SO(p,q)_{0}$. Also Let
\\ $\ Q_{+r} = \{ x \in R^{n+1} / Q_{p +1, q} ( x , x )= r ^{2}\}
, r \in R ^{+} , p,q \in N , n = p + q \geq 1$. \\ Clearly, $\ O( p +
1, q)$ acts on $\ Q _{+r} $ . Let $\ \{ e_{1}, e_{2},
...e_{n}  \}$ be the standard basis for $\ R^{n} $. Then
we have the following result.

\noindent\textbf{Proposition : } For $\ p , q > 0 $, the group
$\ SO_{0}(p + 1, q)$ acts transitively on $\ Q _{+r} $. The
isotropy sub group at  $\ re_{1}$ is isomorphic to $\ SO_{0}
(p,q)$. As a
manifold,\\
$\  Q _{+r} \simeq  SO_{0} ( p + 1, q) / SO_{0} (p,q)$.\\
In particular for $\ n \geq 2, q = n-1 $ and $\ p = 1 $, we define
the semi algebraic cone $\ C $ in $\ R^{n}$ by \\ $\ C $ = $\ \{v \in R^{n
}/ Q_{1,q}(v, v) \geq 0, x\geq 0\}$ and set \\ $\ C^{*} = \Omega =
\{v \in R^{n} / Q_{1,q}(v, v)
> 0, x > 0 \}$. $\ C $ is called the \emph{forward light cone}
in $\ R^{n}$. We have
\noindent $\ M = \left(
\begin{array}{r}
x \\
y
\end{array}
\right) \in C $  if and only if $\   x \geq
 \parallel y \parallel $. \\
(Gheorghe and Mihul \cite{GM} state in Lemma 1 that \emph{There is
a norm $\parallel \parallel$ in $ \overline{M}$ ( a n-1
dimensional linear real space) so that: $\ Q = \{x/ x\in M;
\varepsilon x^{0} = \parallel \overline{x} \parallel\}, intC =
\{x/ x\in M;\varepsilon x^{0} > \parallel \overline{x}
\parallel\}$, where $\varepsilon = 1 $ if $\ (-1,\overline{0})$ is
not in $\ C $ and $\varepsilon = -1 $ if $\ (1,\overline{0})$ is not in
$\ C $)}.

\noindent Boundary of $\ C $ and $\ C^{o} $   are described  as
follows: $\ \partial C  =   \{ v \in R^{n} / \epsilon x =
\parallel y
\parallel \}, \ C^{o} = \{v \in R^{n} / \epsilon x > \parallel y \parallel \} $
where $\ \epsilon $ = 1 if $\ (-1, 0) $ is not in  $\ C $ and $\
\epsilon \neq 1 $ if $\  (1, 0) \in C $.
\\ If  $\ v \in C \cap -C $,  then $\ 0 \leq x \leq 0 $ and hence $\
x $ = 0. Then $\ \parallel y \parallel = 0 $ and thus $\ y = 0$.
Thus $\ v = 0$  and $\ C $ is proper. \\ For $\ v, v' \in C $, we
calculate
\\ $\ (v, v') = (v',v) = x^{'}x + (y^{'} , y) \geq \parallel y'
\parallel
\parallel  y \parallel + (y',y) \geq 0$. Thus $\ C \subset C^{*}$.

\noindent Conversely, let $\  v = \left(
\begin{array}{r}
x \\
y
\end{array}
\right) \in C^{*}$. Then testing against $\ e_{1}$, we get $\ x
\geq 0$. We may assume $\ y \neq 0 $. Define $\ \omega $  by $\
p_{r_{1}} (\omega) = \parallel y \parallel $ and $\ p_{r_{2}}
(\omega)  = -y$. Then $\ \omega \in C $ and  $\ 0 \leq (w, v) = x
\parallel y \parallel - \parallel y
\parallel^{2} = (x - \parallel y \parallel) \parallel y \parallel
$. Hence $\ x \geq \parallel y \parallel$. Therefore $\ y \in C $
and thus $\ C^{*} \subset C$. So $\ C = C^{* }$ and $\ C $ is
self-dual. Similarly, we can show that $\ \Omega $ is self dual.

\noindent Moreover, the forward light cone $\ C $ is invariant under the usual
operation of $\ S O_{o}(1,q)$ and under all dilations, $\ \lambda
I_{n},  \lambda > 0$. ($\ I_{n}$  is the $\ n \times \ n $
identity matrix). We now prove that the group $\ S O_{o}(1,q)
R^{+} I_{q+1}$ acts transitively on $\ \Omega = C^{o} $ if $\ q
\geq 2 $ ( $\ q = 3$ for Minkowski space). Thus $\ \Omega $ will be
homogeneous.For this we prove that $\ \Omega  = SO_{o} (1,q) R^{+} \left(
\begin{array}{r}
1 \\
0
\end{array}
\right)$. \\
Using  \\
$\ a_{t} = \left(
\begin{array}{rrr}
cosh(t) & sinh(t) & 0 \\
sinh(t) & cosh(t) & 0 \\
0 & 0 & I_{n-2}
\end{array}
\right) \in SO_{o} (1,q)$,  we get \\
$\ a_{t}\left(
\begin{array}{r}
\lambda \\
0
\end{array}
\right) = \lambda^{t} (cosh(t), sinh(t), 0, \cdots ,0) $ for all
$\ t \in R$. Let $\ S^{q-1}$ denote a unit sphere in $\ R^{q}$.
Now $\ SO(q)$ acts transitively on $\ S^{q-1} $ and $\ \left(
\begin{array}{rr}
1 & 0 \\
0 & A
\end{array}
\right) \in S O_{o}(1,q)$ \\ for all $\ A \in SO(q)$. Hence the
result follows by noting the fact that coth(t) runs through $\
(1,\infty)$ as t varies in $\ (0,\infty)$.

\noindent There is a vast literature on homogeneous convex cones and they are used in convex
optimization problems. See, for example, the paper by Truong and Tuncel \cite{Tru} and references therein.

\section{Causal Structure of Space-times,Causality Conditions and Causal group}

In this section, we begin with basic definitions and properties of
causal structure of space-time. Then we define different causality
conditions and their hierarchy. Furthermore we discuss causal
group and causal topology on space-time in general, and treat\cite{}
Minkowski space as a special case. We take a space-time ($\ M $,
g) as a connected $\ C^{2} $ - Hausdorff  four dimensional
differentiable manifold which is paracompact and admits a
Lorentzian  metric  g of signature (-, +, +, + ). Moreover, we
assume that the space-time is space and time oriented.

\noindent We say that an event $\ x $ \emph{chronologically} precedes
another event $\ y $, denoted by $\ x \ \ll \  y $ if there is a
smooth future directed  timelike curve from $\ x $ to $\ y $ . If
such a curve is non-spacelike, i.e., timelike or null , we say
that $\ x $ causally precedes $\ y $  or $\ x < \  y $. The
chronological future $\ I^{+}(x) $ of $\ x$ is the set of all
points $\ y$ such that $\ x \ \ll \ y $ . The chronological past
$\ I^{-}(x) $ of $\ x$ is defined dually. Thus

     $\ I^{+}(x)  = \{  y \in \ M /  x \ \ll \  y \} $ \ and

     $\ I^{-}(x)  = \{  y \in \ M /  y \ \ll \  x \} $.

\noindent The causal future and causal past for $\ x $  are defined
similarly :

    $\ J^{+}(x)  = \{  y \in \ M /  x \ < \  y \} $ \ and

      $\ J^{-}(x)  = \{  y \in \ M /  y \ < \  x \} $

\noindent As Penrose \cite{Pen} has proved, the relations $\ \ll $  and $\ < $ are
transitive. Moreover,\\ $\  x \ll \ y $ and $\ y < z $ or $\ x < y
$ and $\ y \ll z $ implies $\ x \ll z $. Thus  $\ \overline
{I^{+}(x)} = \ \overline {J^{+}(x)} $   and also $\  \partial
I^{+}(x) = \ \partial J^{+}(x) $, where for a set $\ X \subset \ M
$, $\ \overline{X} $ denotes closure of $\ X $ and  $\ \partial X
$ denotes topological boundary of $\ X $. The chronological future
and causal future of any set $\ X \subset \  M $ is defined as

    $\ I^{+}(X) = \displaystyle\bigcup_{ x \in  X } I^{+}(x) $ and

    $\   J^{+}(X) = \displaystyle\bigcup_{ x \in  X } J^{+}(x) $ \\
The chronological and causal pasts for subsets of $\ M $ are
defined similarly.

\noindent An ordering which is reflexive and transitive is called
quasi - ordering. This ordering was developed in a generalized
sense by Sorkin and Woolgar \cite{SW} and these concepts were
further developed by Garcia Parrado and Senovilla \cite{GPS1,GPS2}
and S. Janardhan and Saraykar \cite{JS1} to prove many interesting
results in causal structure theory in GR.

\noindent In the recent paper, Zapata and Kreinovich \cite{ZK}  call
chronological order as open order and causal order as closed order
and prove that under reasonable assumptions, one can uniquely
reconstruct an open order if one knows the corresponding closed
order. For special theory of relativity, this part is true and
hence every one-one transformation preserving a closed order
preserves open order and topology. This fact in turn implies that
every order preserving transformation is linear. The conserve part
is well known namely, the open relation uniquely determines both
the topology and the closed order.

\noindent We now introduce the concept of K-causality and give causal properties
of space-times in the light of this concept. For more details we refer the reader to
\cite{JS1}, \cite{Mi1,Mi2}  and \cite{GPS1,GPS2}.

\noindent\textbf{Definition :} \emph{$\ K^{+}$} is the smallest
relation containing $\ I^{+}$ that is topologically closed  and
transitive. If  $\ q $ is in $\ K^{+}(p)  $ then we write $\ p \prec q
$.

\noindent That is, we define the relation $\ K^{+}$, regarded as a subset of
$\ M \times M $, to be the intersection of all closed subsets $\ R
\supseteq I^{+} $ with the property that $\ (p, q)  \in  \ R $ and
$\ (q, r) \in \ R $   implies $\ (p, r) \in \ R $. ( Such sets R
exist because $\ M \times M $ is one of them.) One can also
describe $\ K^{+}$ as the closed-transitive relation generated by
$\ I^{+}$.

\noindent\textbf{Definition :}\ An open set O is \emph{K-causal}
iff the relation `$\ \prec $' \ induces a reflexive partial
ordering on O. i.e. $\ p \prec q $ and $\ q \prec p $ together
imply $\ p = q $.

\noindent If we regard $\ C^{o} $ as the interior of future light
cone in a Minkowski space-time ($\ p = 1, q = 3$ ), then under
standard chronological structure $\ I^{+} , M (a,b)$ becomes $\
I^{-}(b) \cap I^{+}(a)$. As it is well known, such sets form a
base for Alexandrov topology and since Minkowski space-time is
globally hyperbolic and hence strongly causal, Alexandrov topology
coincides with the manifold topology (Euclidean topology). Thus,
lemma 2 of \cite{GM} is a familiar result in the language of
Causal structure theory.

\noindent Analogous to usual causal structure, we  defined in \cite{JS1} strongly
causal and future distinguishing space-times with respect to $\
K^{+}$ relation.

\noindent\textbf{Definition :} \  A $\ C^{0} $ - space-time $\ M $
is said to be \emph{strongly causal at $\ p $ with respect to $\
K^{+}$}, if $\ p $ has arbitrarily small K - convex open
neighbourhoods. \\
Analogous definition would follow for $\ K^{-} $.\\
$\ M $ is said to be \emph{strongly causal with respect to $\
K^{+}$}, if it is strongly causal with respect to $\ K^{+}$ at
each and every point of it. Thus, lemma 16 of \cite{SW} implies that
K-causality implies strong causality with respect to $\ K^{+} $.

\noindent\textbf{Definition :} \ A $\ C^{0} $-  space-time $\ M $
is said to be \emph{K-future distinguishing} if for every $\ p
\neq q , K^{+}(p) \neq K^{+}(q) $. \ \emph{K-past distinguishing}
spaces can be defined analogously.

\noindent\textbf{Definition :} \ A $\ C^{0} $- space-time $\ M $
is said to be \emph{K-distinguishing }if it is both K-future and
K-past distinguishing.

\noindent Analogous result would follow for $\ K^{-} $. Hence, in  a $\
C^{0}$ - space-time  $\ M $, strong causality with respect to K
implies K-distinguishing.

\noindent\textbf{Remark :} \ K-conformal maps preserve K-
distinguishing,  strongly causal with respect to $\ K^{+}$ and
globally hyperbolic properties.

\noindent\textbf{Definition :} \ A $\ C^{0} $- space-time $\ M $
is said to be \emph{K-reflecting} if  \\ $\ K^{+}(p) \supseteq
K^{+}(q) \Leftrightarrow \  K^{-}(q) \supseteq K^{-}(p) $.\\
However, since the condition $\ K^{+}(p) \supseteq  K^{+}(q) $
always implies $\ K^{-}(q) \supseteq K^{-}(p) $ because of
transitivity and $\ x \in K^{+}(x) $, and vice versa, a $\ C^{0}$
- space-time  with K-causal condition is always K-reflecting.
Moreover, in general, K-reflecting need not imply reflecting.
Since, any K-causal space-time is K-reflecting, any non-reflecting
open subset of the space-time will be K-causal but non-reflecting.

\noindent We now give the interesting hierarchy of K-causality
conditions as follows: \\ We have proved that strong causality
with respect to $\ K^{+} $ implies K-future distinguishing. Thus,
K-causality $\ \Rightarrow $ \ strongly causality with respect to
K  $\ \Rightarrow $ \ K - distinguishing.

\noindent Since  a K- causal space-time is always K-reflecting, it follows
that the K-causal space-time is K-reflecting as well as
K-distinguishing. In the classical causal theory, such a
space-time is called causally continuous \cite{HS}. (Such space-times
have been useful in the study of topology change in quantum
gravity \cite{DGS}). Thus if we define K-causally continuous space-time
analogously then we get the result that a K-causal $\ C^{0}$ -
space-time   is K-causally continuous. Moreover, since $\
K^{\pm}(x) $ are topologically closed by definition, analogue of
causal simplicity is redundant and causal continuity (which is
implied by causal simplicity) follows from K-causality. In  \cite{Mi1},
E. Minguzzi proved the equivalence of K-causality and stable
causality.
\\ Thus  the causal hierarchy  reads as follows.

\noindent Global hyperbolicity $\ \Rightarrow $ Stably causal $\
\Leftrightarrow $ K-causality  $\ \Rightarrow $ Strong causality
\\
$\ \Rightarrow $ K - Distinguishing.

\noindent We now proceed to discuss causal groups and causal topology.
We then compare these notions with those in section 2.\\
If $\ R^{n}$ is a directed set with respect to a certain partial
ordering relation `$\ \geq $' of $\ R^{n}$, then such a relation
is called a \emph{Causal relation}. Thus in a globally hyperbolic
space-time (or in a Minkowski space time) $\ J^{+}$ and $\ K^{+}$
are causal relations (In a $\ C^{2}$ globally hyperbolic
space-time, $\ J^{+} = K^{+}$, whereas in a $\ C^{0}$ - globally
hyperbolic space-time, only $\ K^{+}$ is valid). The \emph{Causal
group} $\ G$ relative to causal relation is then defined as the group
of permutations $\ f : R^{n} \rightarrow R^{n}$ which leaves the
relation `$\ \geq $' invariant. i.e. $\ f(x) \geq f(y) $ if and
only if $\ x \geq y $. Such maps are called causal maps. They
preserve causal order. These maps are special cases of  cone
preserving maps defined in section 2.

\noindent Thus in a $\ C^{0}$ globally hyperbolic space-time,
every K - causal map $\ f $ where $\ f^{-1} $ is also order
preserving is a causal relation and causal group is the group of
all such mapping which we called K - conformal groups.

\noindent In the light of the definition of quasiorder given in
section 2, we observe that causal cones and K - causal cones fall in
this category, since causal relation `$\ < $ ' and  K - causal
relation `$\ \prec $ ' are reflexive and transitive. If we replace
quasi-order by a causal relation ( or K-causal relation), then we
see that an order preserving map is  nothing but a causal map.
Thus an order preserving map is a generalization  of a causal map
( or K-causal map).  These concepts also appear in a branch of
theoretical computer science called domain theory. Martin and
Panangaden  \cite{MP} and  S. Janardhan and Saraykar  \cite{JS2}
have used these concepts in an abstract setting and proved some
interesting results in causal structure of space times. They
proved that order gives rise to a topological structure.

\noindent As far as the \emph{causal topology} on $\ R^{n}$ is
concerned, it is defined as the topology generated by the
fundamental system of neighbourhoods containing open ordered sets\\
$\ M(a,b)$ defined for any $\ a, b \in R^{n} $  with $\ b - a \in
C^{o} $ as : $\ M(a,b) = \{ y \in R^{n} / b - y, \ y - a \in
C^{o}\} $. Gheorghe and Mihul \cite{GM} describe  \emph{`causal
topology'} on $\ R^{n}$ and prove that the causal topology of $\
R^{n}$ is equivalent to the Euclidean topology. Causal group is
thus comparable to conformal group of space-time under
consideration. Further any $\ f \in G $ is a homeomorphism in
causal topology and hence it is a homeomorphism in Euclidean
topology.

\noindent If C is a Minkowski cone as discussed in the above
example, then Zeeman \cite{Z} has proved that $\ G$ is generated by
translations, dilations and orthochronous Lorentz transformations
of Minkowski
space $\ R^{n} \ (n = 4)$. \\
We can say more for the causal group $\ G$ of Minkowski space.
\\ Let $\ G_{0} = \{ f \in G/ f(0)= 0\} $ .\\
Then $\ G_{0}$ contains the identity homeomorphism. Gheorghe and
Mihul \cite{GM}  proved that $\ G$ is generated by the translations of $\
R^{n} $ and by linear transformation belonging to $\ G_{0}$. Hence
$\ G$ is a subgroup of the affine group of $\ R^{n}$. This is the main
result of \cite{GM}.

\noindent Let $\  G^{'}_{0} =  G_{0} \cap SL(n, R)$. Then $\
G^{'}_{0}$ is the orthochronus Lorentz group under the norm $\
\parallel y
\parallel = [\displaystyle\sum_{i=1}^{q} \mid y^{i} \mid^{2}]^{\frac{1}{2}}$
 for $\ y \in R^{q}, y = (y^{1}, y^{2},\cdots,y^{q}) $.\\
\hspace{10mm} For $\ \parallel y \parallel =
[\displaystyle\sum_{i=1}^{q} \mid y^{i}
\mid^{\alpha}]^{\frac{1}{\alpha}}$,  $\ \alpha > 2 , \ G^{'}_{0}$
is the discrete group of permutations and the symmetries relative
to the origin of the basis vectors of $\ R^{q}$. The factor group
$\ G_{0}/G_{0}^{'}$ is the dilation group of $\ R^{n}$. Also, $\ G$ is
the semi-direct product of the translation group with the subgroup
$\  G^{'}_{0}$  of SL(n,R). Moreover $\  G^{'}_{0}$ is a
topological subgroup of SL(n,R). Similar results have been proved by Borchers and
Hegerfeldt \cite{BH}. Thus we have,

\noindent\textbf{Theorem :} Let $\ M $ denote n-dimensional
Minkowski space, $\ n \geq 3$ and let $\ T $ be a 1 - 1 map of $\ M $
onto $\ M $. Then $\ T $ and $\ T^{-1} $ preserve the relation $\ (x -
y)^{2} > 0 $ if and only if they preserve the relation $\ (x -
y)^{2} = 0 $. The group of all
such maps is generated by \\
i)  The full Lorentz group (including time reversal)\\
ii)Translations of $\ M $ \\ iii) Dilations ( multiplication by a
scalar)
\\ In our terminology, $\ T $ is a causal map. \\ In the same
paper \cite{BH}, the following theorem is also proved.

\noindent\textbf{Theorem :} Let $\ dim M \geq 3$, and let $\ T $
be a 1 - 1 map of $\ M $ onto $\ M $, which maps light like lines
onto (arbitrary) straight lines. Then $\ T $ is linear.

\noindent This implies that constancy of light velocity c alone implies the
Poincare group upto dilations.

\noindent Thus, for Minkowski space, things are much simpler. For a
space-time of general relativity (a Lorentz manifold) these
notions take a more complicated form where partial orders are $\
J^{+} $ or $\ K^{+}$ .

\section{Causal Orientations and order theoretic approach to Global Hyperbolicity }
In this section, we discuss briefly the concepts of \emph{Causal Orientations ,
Causal Structures and Causal Intervals} which lead to the
definition of a \emph{`Globally hyperbolic homogeneous space'}.\\
These notions cover Minkowski Space and homogeneous cosmological models in general relativity.
We also discuss domain theoretic approach to causal structure of space-time and comment
on the parallel concepts appearing in these approaches.

\noindent Let $\ M $ be a $\ C^{1}$ (respectively smooth) space-time.
For $\ m \in M, T_{m}(M) $ denotes the tangent space of $\ M $ at m, and T($\ M $) denotes
the tangent bundle of $\ M $. The derivative of a differentiate
map $\ f : M \rightarrow N $ at m will be denoted by $\ d_{m}f :
T_{m}M \rightarrow T_{f(m)}N $. A $\ C^{1} $ (respectively smooth)
causal structure on $\ M $ is a map which assigns to each point $\
m \in M $ a nontrivial closed convex cone C(m) in $\ T_{m}M $ and
it is $\ C^{1}$(smooth) in the following sense:

\noindent We can find an open covering \\
$\ \{ U_{i} \}_{i \in I} $ of $\ M $,
smooth maps $\ \phi_{i}: U_{i} \times R^{n} \rightarrow T(M) $with
$\ \phi_{i}(m,M) \in T_{m}(M)$ and a cone C in $\ R^{n} $ such
that $\ C(m)  = \phi_{i} (m,C)$ .

\noindent The causal structure is called \emph{generating} (respectively
proper, regular) if C(m) is generating (proper, regular) for all
m. A map $\ f : M \rightarrow M $ is called  \emph{causal} if $\
d_{m}f(C(m)) \subset C(f(m)) $ for all $\ m \in M $. These
definitions are obeyed by causal structure $\ J^{+}$ in a causally
simple space-time and causal maps of Garc$\acute{i}$a-Parrado and
Senovilla \cite{GPS2}. If we consider $\ C^{0}$- Lorentzian manifold with
a $\ C^{1}$ -metric so that we can define null cones, then
these definitions are also satisfied by causal structure $\ K^{+}$
and K-causal maps. Thus the notions
defined above are more general than those occurring in general
relativity at least in a special class of space-times.   Rainer
\cite{Ra} called such a causal structure an   \emph{ultra weak cone
structure}  on $\ M $ where $\ m \in \ int M  $.

\noindent We now define $\ G$- invariant causal structures where $\ G$ is a Lie group and discuss
some properties of such structures. If a Lie group $\ G$ acts smoothly on
$\ M $ via $\ (g,m) \mapsto g.m.$, we denote the diffeomorphism $\ m \mapsto g.m $ by $\
l_{g}$.

\noindent\textbf{Definition:} Let $\ M $ be a manifold with a
causal structure and $\ G$ a Lie group acting on $\ M $. Then the
causal structure is called \emph{$\ G$ - invariant} if all $\ l_{g}, \
g \in G $, are causal maps. If H is a Lie subgroup of $\ G$ and $\ M =
G/H $ is homogeneous then a $\ G$-invariant causal structure is
determined completely by the cone $\ C = C(0) \subset T_{o}M $,
where $\ o = {H} \in G/H $. Moreover C is proper, generating etc
if and only if this holds for the causal structure. We also note
that C is invariant under the action of H on $\ T_{o}(M)$ given by
$\ h \mapsto d_{0}l_{h}$. On the other hand, if $\ C \in Cone_{H}
(T_{o}(M))$,then we can define a field of cones by $\ M
\rightarrow T_{\alpha .0}(M) :
\\ aH \mapsto C(\alpha H) = d_{0} l_{a}(C)$.

\noindent This cone field is $\ G$-invariant, regular and satisfies $\ C(0) = C $.
Moreover the mapping $\ m \mapsto C(m)$ is also smooth in the
sense described above. If this mapping is only continuous in the
topological sense,  for all m  in M,  then Rainer \cite{Ra} calls such
cone structure, a \emph{weak local cone structure} on $\ M
$.\\
We have the following theorem.

\noindent\textbf{Theorem :} Let $\ M = G/H $ be homogeneous.
Then $\ C \mapsto (\alpha H \mapsto d_{0}l_{a}(C)) $ defines a
bijection between $\ Cone_{H}(T_{o}(M)) $ and the set of
$\ G$-invariant, regular causal structures on $\ M $.

\noindent We call  a mapping $\ \nu : [a,b] \rightarrow M $ as
\emph{absolutely continuous }if for any coordinate chart $\ \phi :
U \rightarrow R^{n}$, the curve $\ \eta = \phi \circ \nu :
\nu^{-1}(U) \rightarrow R^{n} $ has absolutely continuous
coordinate functions and the derivatives of these functions are
locally bounded.

\noindent Further, we define a \emph{C-causal curve}: \ Let $\ M = G/H $ and
$\ C \in Cone_{G}(T_{o}M)$. An absolutely continuous curve $\ \nu
: [a,b] \rightarrow M $ is called C - causal ( \emph{Cone causal
or conal}) if $\ \nu^{'}(t) \in C(\nu(t)) $ whenever the
derivative exists.

\noindent Next, we define a relation `$\ \leq_{s}$'  (s for strict) of $\ M
$ by $\ m \leq_{s} n $ if there exists a C-causal curve $\ \nu $
connecting m with n. This relation is obviously reflexive and
transitive. Such relations are called \emph{causal orientations}
or \emph{quasi - orders}. They give rise to causal cones as we saw in section 2.

\noindent\textbf{Note :} A reader who is familiar with the books
by Penrose \cite{Pen}, Hawking and Ellis \cite{H&E} or Joshi \cite{Jo} will
immediately note that the above relation is our familiar causal
order $\ J^{\pm}$ in the case when $\ M $ is a space-time in
general relativity.

\noindent We ask the question : Which of the space-times $\ M $
can be written as $\ G/H $?  G\"{o}del universe, Taub universe and
Bianchi universe are some examples of such space-times. They are
all spatially homogeneous cosmological models. Isometry group of a
spatially homogeneous cosmological model may or may not be
abelian. If it is abelian, then these are of Bianchi type I, under
Bianchi classification of homogeneous cosmological models. Thus
above discussion applies to such models.

\noindent As an example to illustrate above ideas, we again consider a
finite dimensional vector space $\ M $ and let C be a closed
convex cone in $\ M $. Then we define a causal Aut(C) - invariant
orientation on $\ M $ by   $\ u \leq v $ iff $\  v - u \in C $ .
Then `$\ \leq $' is antisymmetric iff C is proper. In particular
$\ H^{+}(n,\emph{R}) $ defines a $\ GL(n,\emph{R})$ -invariant
global ordering in H(n,\emph{R}). Here H(n,R) are $\ n\times n $
real orthogonal matrices (Hermitian if \emph{R} is replaced by
\emph{C}) and  $\ H^{+}(n,R) = \{ X \in H( m,\emph{R}) / X $ is
positive definite $\}$ is an open convex cone in H(n,R). (the
closure of $\ H^{+}(n,R) $ is the closed convex cone of all
positive semi definite matrices in H(n,R)). Also, the light cone
$\ C \subset R^{n+1 }$ defines a $\ SO_{O}(n,1)$ -invariant
ordering in $\ R^{n+1}$. The space $\ R^{n+1}$ together with this
global ordering is  the (n+1)-dimensional Minkowski space.

\noindent Going back to the general situation we note that in general, the
graph \\ $\ M_{\leq_{s}} = \{(m,n) \in M \times M / m \leq_{s} n
\} $ of `$\ \leq_{s}$' is not closed in $\ M \times M $. However,
if we define $\ m \leq n \Leftrightarrow (m,n) \in
\overline{M}_{\leq_{s}}$, then it turns out that `$\ \leq $' is a
causal orientation. This can be seen as follows:\\
`$\ \leq $'  is obviously reflexive. We show that it is
transitive: \\ Suppose $\ m \leq n \leq p $ and let $\
m_{k},n_{k},n_{k}^{'},p_{k} $ be sequences such that $\ m_{k}
\leq_{s} n_{k}, \\ n_{k}^{'} \leq_{s} p_{k}, \ m_{k} \rightarrow \
m,   \ n_{k} \  \rightarrow \ n, \  n_{k}^{'} \ \rightarrow \ n $
and $\ p_{k} \rightarrow  \ p $. Now we can find a sequence $\
g_{k} $ in $\ G$ converging to the identity such that $\ n_{k}^{'} =
g_{k}n_{k}$. Thus $\ g_{k}m_{k} \rightarrow  m $ and $\ g_{k}n_{k}
\leq_{s} p_{k} $ implies $\ m \leq p $.

\noindent The above result resembles the way in which $\ K^{+} $ was
constructed from $\ I^{+}$.\\
The following definitions are analogous to $\ I^{\pm},J^{\pm}$ or
$\ K^{\pm}$ and so is the definition of interval as $\ I^{+}(p)
\cap I^{-}(q) ( J^{+}(p) \cap J^{-}(q)$  or $\ K^{+}(p) \cap
K^{-}(q))$: \\
Given any causal orientation `$\ \leq $' on $\ M $, we define for
$\ A \subset M$,

\noindent $\  \uparrow A = \{ y \in M / \exists a \in A \ $ with $\ a \leq y
\}$   and

\noindent $\  \downarrow A = \{ y \in M / \exists a \in A \ $ with $\ y \leq
a
\}$.\\
Also, we write $\ \uparrow x = \uparrow \{x\} $ and $\ \downarrow
x = \downarrow \{x\} $. \\
The intervals with respect to this causal orientation are defined
as

\noindent $\ [m,n]_{\leq}= \{z \in M / m \leq z \leq n \} = \uparrow m \
\cap
\downarrow n $ .\\
Finally we introduce some more definitions.

\noindent\textbf{Definitions :} Let $\ M $ be a space-time.\\
(1) a causal orientation `$\ \leq $' on $\ M $ is called
\emph{topological} if its graph $\ M_{\leq} $ in $\ M \times M $
is closed. \\
(2) a space $\ (M,\leq )$  with a topological causal orientation
is called a \emph{causal space}. If `$\ \leq $' is, in addition,
antisymmetric, that is a partial order, then $\ (M,\leq )$ is
called \emph{globally ordered} or \emph{ordered}.\\
(3) Let $\ (M, \leq )$ and $\ (N,\leq )$ be two causal spaces and
let $\ f:  M \rightarrow  N $ be continuous. Then $\ f $ is called
\emph{order preserving} or \emph{monotone} if $\ m_{1} \leq  m_{2}
 \Rightarrow f(m_{1}) \leq f(m_{2})$. \\
(4) Let $\ G$ be a group acting on $\ M $. Then a causal orientation
$\ \leq $ is called \emph{$\ G$-invariant} if $\ m \leq n  \Rightarrow
a.m \leq  a.n , \  \forall \ a \in G $.\\
(5) A triple $\ ( M, \leq, G)$ is called a \emph{Causal
$\ G$-Manifold} or \emph{causal} if `$\ \leq $' is a topological
$\ G$-invariant causal orientation.
\\ Thus referring to partial order $\ K^{+}$, we see, in the light of
above definitions (1) and (2), that $\ \leq_{K}$ is topological
and $\ ( M, \leq_{K})$ is a causal space. A K-causal map satisfies
definition (3).

\noindent For a homogeneous space $\ M = G/H $ carrying a causal
orientation such that $\ ( M, \leq, G)$ is causal, the intervals
are always closed subsets of $\ M $. If the intervals are compact,
we say that $\ M = G/H $ is \emph{globally hyperbolic}. We use the
same definition for a space-time where intervals are $\
J^{+}(p)\cap J^{-}(q)$. Thus globally hyperbolic space-times can
be defined by using causal orientations for homogeneous spaces. In
this setting, intervals are always closed, as in causally
continuous space-times.

\noindent As the last part of our review, we discuss the central concepts and definitions
of domain theory, as we observe that these concepts  are related to causal
structure of space-time  and also to space-time topologies.

\noindent The relations $\ < and \ll $ discussed in section 3 have been generalised
to abstract orderings using the concepts in Domain Theory and also many
interesting results have been proved related to causal structures of space - time in
general relativity. For definitions and preliminary results in domain theory,
we follow Abramsky and Jung \cite{AbJ} and Martin and Panangaden
\cite{MP}.

\noindent A \emph{poset} is defined as a partially ordered set, i.e. a set
together with a reflexive, anti- symmetric and transitive
relation.A concept that plays an important role in the theory is the one of
a directed subset of a domain, i.e. of a non-empty subset  in
which each two elements have an upper bound.

\noindent\textbf{Definition : } Let $\ (P, \sqsubseteq)$ be a
partially ordered set. An  \emph{upper bound}  of a subset $\ S $ of a
poset $\ P $ is an element b of $\ P $, such that $\ x \sqsubseteq  b , \
\forall x \in \ S $. The dual notion is called \emph{lower bound}.

\noindent\textbf{Definition : }A nonempty subset $\ S \subseteq P$
is \emph{directed} if for every $\ x, y $ in S, $\ \exists \ z \in
S \  \ni: \ x , y \sqsubseteq \ z $. The supremum of $\ S
\subseteq P$ is the least of all its upper bounds provided it
exists and is denoted by $\ \bigsqcup S $.

\noindent A nonempty subset $\ S \subseteq P$ is  \emph{filtered } if for
every $\ x $, y in S ,$\ \exists z \in S \ni: \ z  \sqsubseteq \ x
, y $. The  infimum of $\ S \subseteq P$ is the greatest of all
its lower bounds provided it exists and is denoted by $\ \bigwedge
S $.

\noindent \textbf{Definition :} A \emph{dcpo} is a poset in which
every directed subset has a supremum.\\
Using partial order some topologies can be derived. For example,

\noindent\textbf{Definition :} A subset U of a poset $\ P $ is
 \emph{Scott open} if \\
(i) U is an upper set: i.e. $\ x \in U $ and $\  x \sqsubseteq y
 \Rightarrow  y \in U $ and \\
(ii) U is inaccessible by directed suprema: i.e. for every
directed $\ S \subseteq P $ with a supremum,
$\ \bigsqcup S \in U \Rightarrow  S \cap U \neq \phi $. \\
The collection of all Scott open sets on $\ P $ is called the
\emph{Scott topology}.
\\ The order of \emph{approximation}  denoted by $\
`\ll '$ is defined as:

\noindent\textbf{Definition :}  For elements $\ x , y $ of a
poset, $\ x \ll y $  iff for all directed sets S with a supremum,
$\ y  \ \sqsubseteq  \bigsqcup S \ \Rightarrow \exists \ s \in S
\ni: \ x \sqsubseteq S $. \\
Define, $\ \Downarrow x = \{ a \in  D / a \ll x  \}$  and $\
\Uparrow x = \{ a \in  D /  x \ll a \}$ .

\noindent The special property of the finite elements $\ x $ is that they
are way-below themselves, i.e. $\ x \ll x $. An element with this
property is also called \emph{compact}.

\noindent\textbf{Definition :} For a subset X of a poset $\ P $, define
\\ $\ \uparrow X := \{y \in P / \exists \ x \in X,  x
\sqsubseteq y \}$  and \\
$\ \downarrow X :=  \{y \in P / \exists \ x \in X,  y \sqsubseteq
x \}$ \\
Then, $\ \uparrow x =\uparrow \{x \}$ and $\ \downarrow x =
\downarrow \{x \} $ for $\ x \in X $.

\noindent\textbf{Definition :} A \emph{basis} for a poset D is a
subset B such that $\ B \cap \Downarrow x $ contains a directed
set with supremum $\ x $ for all $\ x $ in  D.
A poset is  continuous if it has a basis. A poset is $\ \omega
$-continuous if it has a countable basis.\\
Then we have,

\noindent\textbf{Theorem:} The collection $\ \{ \Uparrow x /
x \in D \}$ is a basis for the Scott topology on a continuous
poset.
\\ Lawson topology is defined as,

\noindent\textbf{Definition :} \ The \emph{Lawson topology} on a
continuous poset $\ P $ has as a basis all sets of the form $\ \Uparrow
x \sim \uparrow F $, for $\ F \subseteq P $ finite.

\noindent\textbf{Definition : } A continuous poset $\ P $ is
\emph{bicontinuous} if for all $\ x , y $ in  $\ P $\\
$\ x  \ll y $ iff for all filtered $\ S \subseteq P $ with an
infimum, $\ \bigwedge S \sqsubseteq x \Rightarrow \exists \ s \in
\ S \ni:  \ s \sqsubseteq \ y $ and   for each $\ x \in P $, the
set $\ \Uparrow x $ is filtered with infimum $\ x $.

\noindent\textbf{Theorem: } On a bicontinuous poset $\ P $, sets of
the form $\ (a, b) := \{ x \in P / a \ll x \ll b \} $  form a basis for a
topology.
This topology is called the  \emph{interval topology}.

\noindent\textbf{Definition :} \ The Alexandrov topology on a
space-time has $\{I^{+}(p)\cap I^{-}(q) / \ p, q \in \ M \}$ as a
basis.
\\ For a pre-ordered set $\ P $, any upper
set O is Alexandrov-open. Inversely, a topology is Alexandrov if
any intersection of open sets is open.

\noindent Let $\ I^{\pm}(p) $ and $\ J^{\pm}(p)$ be defined as in section 3.
The relation $\ J^{+} $ can be defined \\ as $\ p \sqsubseteq q
\equiv q \in J^{+}(p) $. Then we have the following:

\noindent\textbf{Proposition :}  Let $\ p, q, r \in M $. Then \\
(i) The sets $\ I^{+}(p) $ and $\ I^{-}(p) $ are open. \\
(ii) $\ p \sqsubseteq q $ and $\ r \in I^{+}(q) \Rightarrow  r \in
I^{+}(p)$ \\
(iii)$\  q \in  I^{+}(p)$ and $\ q \sqsubseteq r \Rightarrow r \in
I^{+}(p)$ \\
(iv) $\ \overline{I^{+}(p)} = \overline{J^{+}(p)}$ and $\
\overline{I^{-}(p)} = \overline{J^{-}(p)}$.

\noindent\textbf{Theorem :} A space-time $\ M $ is strongly
causal iff its Alexandrov topology is Hausdorff iff its Alexandrov
topology is the manifold topology.

\noindent\textbf{Proposition :} $\ x \ll y  \Leftrightarrow y
\in I^{+}(x)$  for all $\ x , y $ in $\ M $.

\noindent\textbf{Theorem :}  $\ (M, \sqsubseteq ) $ is a
bicontinuous poset with $\ \ll = I^{+}$ whose interval topology is
the manifold topology.
\\ Causal simplicity  also has a characterization in order-theoretic
terms.

\noindent\textbf{Theorem :} Let   $\ (M, \sqsubseteq) $ be a
continuous poset with $\ \sqsubseteq = I^{+}$. Then the following
are
equivalent: \\
(i) $\ M $ is causally simple. \\
(ii) The Lawson topology on $\ M $ is a subset of the interval
topology on $\ M $.

\noindent \textbf{Definition :} A \emph{globally hyperbolic}
space-time $\ (M, \sqsubseteq ) $ is a bicontinuous poset whose
intervals are compact in the interval topology on M.\\
Bicontinuity ensures that the topology of M, that is, the interval
topology is implicit in $\ \sqsubseteq $.

\noindent \textbf{Theorem:} \ A globally hyperbolic poset is
locally compact Hausdorff. Also,\\
(i) The Lawson topology is contained in the interval topology.\\
(ii) Its partial order $\ \sqsubseteq $ is a closed subset of $\
M^{2}$. \\
We extended and generalized in [9], some of the above
concepts to K- causality in a $\ C^{0} $- globally hyperbolic
space-time as follows.

\noindent \textbf{Result:} \ In a $\ C^{0} $ - globally
hyperbolic space-times, $\ x  \ll y \ \Rightarrow \  y \in
K^{+}(x) $ where the partial order is $\ \prec = K^{+}   $.

\noindent It must be noted that above analysis does not require any kind of
differentiability conditions on a space-time manifold, and results
are purely topological and order- theoretic.
\\ We illustrate, for Lawson topology, as to how the concepts above
can be generalized to K- causality.\\ We also have analogous to above,\\
$\ \Downarrow x = \{ a \in  M  /  a \ll x  \}$ and \\
$\ \Uparrow x = \{ a \in  M  /  x \ll a \}$ .\\
Since $\ a \ll x \Rightarrow a \in K^{-}(x) $, we have, \\
$\ \Downarrow x \subseteq K^{-}(x) $ and $\ \Uparrow x \subseteq
K^{+}(x) $.
\\ Let us now take a basis for Lawson topology as the sets of the
form \\
$\{  \Uparrow x \sim \uparrow F  \ / \ F $ is finite $\}$.

\noindent Since F is finite, F is compact in the manifold topology and hence
$\ \uparrow F $ is closed. Since the sets $\ \Downarrow x $ and $\
\Uparrow x $ are open in the manifold topology ( in a $\ C^{0} $ -
globally hyperbolic space-time ), $\ \Uparrow x \sim \uparrow F $
are also open in the manifold topology .

\noindent Thus Lawson open sets are open  in the manifold topology also and
hence Lawson topology, in K- sense, is contained  in the manifold
topology.

\noindent Similar analysis can be given for Scott topology and interval
topology also. The intervals defined above, with appropriate cone
structure coincide with causal intervals and hence so does the definition of global hyperbolicity. When the
partial order is $\ J^{+} $, interval topology coincides with
Alexandrov topology and as is well-known, in a strongly causal
space-time, Alexandrov topology coincides with the manifold
topology.

\section{ Concluding Remarks}
We note that there are a large number of space-times
(solutions of Einstein field equations) which are inhomogeneous
(see Krasinski \cite{Kra}) and hence do not fall in the above class :
$\ M = G/H $.

\noindent M.Rainer \cite{Ra} defines yet another partial order
using cones as subsets of a topological manifold and a
differential manifold (space-time) which is a causal relation in
the sense defined above and which is more general than $\ J^{+}$.
Rainer, furthermore defines analogous causal hierarchy like in the
classical causal structure theory. Of course, for Minkowski space,
the old and new definitions coincide. For a $\ C^{2}$-globally
hyperbolic space-time $\ J^{+}, K^{+}$  and Rainer's relation all
coincide, whereas for a $\ C^{0}$-globally hyperbolic space-time,
$\ K^{+}$ and Rainer's relation on topological manifold coincide.

\noindent Moreover if the cones are characteristic surfaces of the
Lorentzian metric, then all his definitions of causal hierarchy
coincide with the classical definitions. (cf theorem 2 of Rainer
\cite{Ra}). For more details on this partial order, we refer the
reader to this paper.

\noindent B.Carter \cite{Car} discusses causal relations from a
different perspective and discusses in detail many features of
this relationship. Topological considerations in the light of
time-ordering have been discussed by E.H.Kronheimer \cite{Kro}.

\end{document}